\documentclass{iopart}
\usepackage{ulem}
\usepackage{array}
\usepackage{bbm}
\usepackage{graphicx}
\usepackage{verbatim}
\usepackage{xfrac}
\usepackage{cite}
\usepackage{subfig}
\usepackage{color}

\renewcommand\emph[1]{\textit{#1}}
\renewcommand\em\it

\newcommand\rawCAFE[1]{CAFE$^{(#1)}$}
\newcommand\genCAFE[2]{\rawCAFE{#1}$\times{#2}$}
\newcommand\CAFE[2]{\genCAFE{3,5,#1}{#2}}
\newcommand\rawCRUDD{RUDD$^*$}
\newcommand\CRUDD[2]{\rawCRUDD-$#1$-$#2$}
\renewcommand\vec[1]{\mathbf{#1}}
\newcommand\be{\begin{equation}}
\newcommand\ee{\end{equation}}
\newcommand\ts[2]{#1_{\text{#2}}}
\newcommand\ket[1]{|#1\rangle}
\newcommand\refeq[1]{(\ref{#1})}
\newcommand\reffig[1]{figure~\ref{#1}}
\newcommand\id{\mathbbm{1}}
\newcommand{\vecalphamag}{A}
\newcommand{\vecalpha}{\vec{\vecalphamag}}

\newcommand{\magcon}{\beta}
\newcommand{\dircon}{\hat{\vec{n}}}

\newcommand{\magcont}{\magcon(t)}
\newcommand{\dircont}{\dircon}
\newcommand{\infid}{\mathcal{I}}

\newcommand\titlestuff{
\title{Dynamical decoupling of a qubit with always-on control fields}
\author{N. Cody Jones,
        Thaddeus D. Ladd$^*$, and
        Bryan H. Fong}
\address{HRL Laboratories, LLC., 3011 Malibu Canyon Rd., Malibu, California 90265, USA}

\address{$^{*}$
         Corresponding author:}
         \ead{tdladd@hrl.com}
}

\begin{document}
\titlestuff
\begin{abstract}
We consider dynamical decoupling schemes in which the qubit is continuously manipulated by a control field at all times.  Building on the theory of the Uhrig Dynamical Decoupling sequence (UDD) and its connections to Chebyshev polynomials, we derive a method of always-on control by expressing the UDD control field as a Fourier series.  We then truncate this series and numerically optimize the series coefficients for decoupling, constructing the CAFE (Chebyshev and Fourier Expansion) sequence.  This approach generates a bounded, continuous control field.  We simulate the decoupling effectiveness of our sequence \emph{vs.} a continuous version of UDD for a qubit coupled to fully-quantum and semi-classical dephasing baths and find comparable performance.  We derive filter functions for continuous-control decoupling sequences, and we assess how robust such sequences are to noise on control fields.  The methods we employ provide a variety of tools to analyze continuous-control dynamical decoupling sequences.
\end{abstract}

\section{Introduction}
    \label{introduction}
Quantum information processing seeks to exploit quantum mechanical systems to store and manipulate information in novel ways.  Typically, however, this quantum behavior exists at the microscopic level, such as a single atom or electron, making control by the experimenter challenging.  Moreover, such a system is never perfectly isolated.  Stray electromagnetic fields, thermal fluctuations, \emph{etc.} can perturb the quantum information carrier, imparting noise into the system. This noise can disrupt the quantum information processor and negate any benefits to operating in a quantum system.

Quantum states can be protected from noise with active control fields.  Before the field of quantum information emerged, a wide variety of techniques for dynamical decoupling (DD) of a spin ensemble were developed in the nuclear magnetic resonance (NMR) community.  The
first discovery was the Hahn spin echo sequence~\cite{Hahn1950}.  By repeating the spin echo sequence several times in succession, one finds the Carr-Purcell
(CP) sequence~\cite{Carr1954}.  This sequence was soon improved by selectively choosing the axes around which one rotates the spin vector, yielding the workhorse
Carr-Purcell-Meiboom-Gill (CPMG) sequence~\cite{Meiboom1958}.  Ever more complicated sequences were developed in ensuing decades, allowing detailed characterization or manipulation of the environments of nuclear spin ensembles, or the compensation against inhomogeneous control fields.

Dynamical decoupling found a renewed purpose in quantum information science.
For example, quantum error correction
systems~\cite{Shor1995,Steane1996,Calderbank1996,Knill2005,Bacon2006,Aliferis2007,Fowler2009} demand an error-per-operation of less than about
1\%~\cite{Knill2005,Wang2011}.  Such low error rates can be difficult to achieve in practice since many candidate physical systems will lose their qubit coherence very quickly owing to environment-induced decoherence; as a result, quantum error correction may fail due to short qubit lifetimes.  Dynamical decoupling schemes offer a relatively simple solution to this problem: by applying a determined sequence of control pulses, one can significantly enhance the lifetime of a qubit.

Early work in dynamical decoupling sought to suppress decoherence by applying a periodic sequence of instantaneous ``bang-bang'' pulses, which periodically flip the state of a qubit and undo the coupling to the bath~\cite{Viola1998,Viola1999,Viola2002,Khodjasteh2007}.
This work was extended to Hamiltonian engineering by using decoupling sequences to selectively enable coupling Hamiltonians~\cite{Viola1999a,Lloyd2001}.  Subsequently, attention turned to using Eulerian graphs to design sequences using bounded control operations that were robust to many types of systematic control errors~\cite{Viola2003}.  This led to the notion of dynamically-corrected gates~\cite{Khodjasteh2009}, which combine techniques from dynamical decoupling and composite pulse sequences from NMR to produce error-suppressing quantum gates.

The idealized ``bang-bang'' control pulse is instantaneous, like the Dirac delta function. However, real physical pulses can only approximate ``bang-bang'' control, because such idealized control would require infinite power.  The implications of a real, continuous-time function can be significant.  Most dynamical decoupling sequences are designed to correct noise errors \emph{between} pulses, but offer no intrinsic protection to errors \emph{during} pulses.  Early studies of always-on dynamical decoupling arose in the problem of dipolar decoupling in solid-state NMR~\cite{Burum81}.  More recently, optimal control approaches to customizing continuous, bounded controls to decouple a specific but arbitrary noise bath were considered in Refs.~\cite{kurizki08,kurizki10}.
Of particular relevance to the present work are approaches that extend optimized bang-bang sequences such as the \textbf{U}hrig \textbf{D}ynamical \textbf{D}ecoupling (UDD) sequence~\cite{Uhrig 07}, which is derived by canceling general noise Hamiltonians order-by-order in perturbation theory.   Extensions of UDD to continuous and bounded control were studied by Uhrig and Pasini~\cite{Pasini2009,Uhrig2010}.  This work introduced pulses that correct errors during their own evolution to finite order in a time-dependent perturbation expansion.  These sequences are termed ``realistic UDD,'' or RUDD.  Subsequent work showed how to produce arbitrary gates decoupled from environment noise using bounded-strength controls~\cite{Khodjasteh2009b,greenbiercuk}, as well as the application of modifications based on optimal control given known system drifts~\cite{grace12}.  Dynamical decoupling with such ``finite-width'' pulses represents efforts to improve existing sequences by accounting for the real continuous-time nature of control fields.

Dynamical decoupling sequences where the control field is always or almost always on pose some unique challenges for sequence engineering and characterization.  Methods for characterization that are exact in the ``bang-bang" limit of DD and approximately correct with short but finite pulse widths require continuous analogs.  In this work, we examine methods to both create and characterize always-on sequences, by considering continuous extensions of the UDD sequence.  We introduce a new DD sequence called CAFE (\textbf{C}hebyshev \textbf{a}nd \textbf{F}ourier \textbf{E}xpansion), and compare it to similar RUDD sequences.

The paper is organized as follows.  Section~\ref{theory} gives a derivation of the theory behind CAFE, presenting it as a continuous extension of the bang-bang UDD sequence. Ultimately, CAFE is derived from a particular heuristic rather than from a fundamental determination of optimality, but this is typical of continuous sequences.     Critical for evaluating the heuristic, then, are methods of characterizing the sequence, and careful choices of sequences for comparison, as discussed in section~\ref{comparison}.  Section~\ref{quantum_bath} tests the CAFE sequence as well as RUDD sequences for performance in decoupling a qubit from a simulated quantum bath, represented by a small dipole-coupled spin lattice.  Section~\ref{filter_function} studies the frequency response of CAFE and compares analytical and simulated filter functions for this sequence and RUDD sequences.  Section~\ref{simulation} simulates the scenario where a qubit is coupled to semiclassical noise and where control fields for decoupling are noisy, again for both CAFE and RUDD sequences.

\section{Derivation of CAFE sequences}
    \label{theory}
We consider a qubit that is exposed to an environment that causes phase errors, or a dephasing bath.  For simplicity, we parameterize our system in terms of Pauli operators $\sigma^j$, so that the phase errors correspond to $\sigma^z$ operators and the control of the qubit is proportional to the $\sigma^x$ operator.  The total system Hamiltonian is
\begin{equation}
\mathcal{H}(t) = \id \otimes B_0 + \sigma^z \otimes B_Z + \alpha(t) \sigma^x \otimes \id,
\label{Hamiltonian}
\end{equation}
where $\id$ refers to identity in its subspace.  In this expression, $\id\otimes B_0$ is a pure bath evolution term, $\sigma^z \otimes B_Z$ is the qubit-bath coupling term, and $\alpha(t) \sigma^x \otimes \id$ is the time-dependent term representing the experimenter's control.  In the following analysis, we find it useful to define the time-integrated control function
\begin{equation}
\label{betadef}
\beta(t) = 2 \int_0^t \alpha(s) ds.
\end{equation}
The quantity $\beta(t)$ has an intuitive meaning --- it represents the total angle (commonly called ``pulse area'') that the qubit state is rotated on the Bloch sphere, in this case around the $X$ axis.  We now transform our system to the interaction picture, so that the interaction Hamiltonian becomes
\begin{equation}
\mathcal{H}_{\mathrm{int}}(t) = \cos\left[\beta(t)\right] \sigma^z \otimes B(t) + \sin\left[\beta(t)\right] \sigma^y \otimes B(t),
\label{H_int}
\end{equation}
where
\begin{equation}
B(t) = e^{iB_0 t} B_Z e^{-iB_0 t}.
\end{equation}

We see that, in the interaction picture, the system-bath coupling terms are functions of $\beta(t)$.  In the case of ``bang-bang'' $\pi$-pulse control, $\beta(t)$ is a piecewise-constant function that is everywhere a multiple of $\pi$, so that $\sin\left[\beta(t)\right] = 0$ everywhere and we can replace $\cos\left[\beta(t)\right]$ with a ``switching function'' $y(t)$ that takes on the values $\pm 1$.  For example, this is the formalism considered in Ref.~\cite{Uhrig07}.  We will later refer to the theory behind the bang-bang version of UDD to derive always-on CAFE sequences.

\subsection{Dynamical Decoupling by Order in $T$}
Our task is to develop a class of control functions $\beta(t)$ so that the system evolution of the qubit over time interval $t \in [0,T]$, given by
\begin{equation}
\mathcal{U}(0,t) = \mathcal{T}\exp\left(-i\int_0^t \mathcal{H}(\tau) d\tau\right),
\end{equation}
closely approximates the identity operation on the qubit (to within a complex phase factor) at the final time $t=T$, where $\mathcal{T}$ is the time-ordering operator.  We approach this problem by expanding the interaction picture propagator in the Magnus expansion, as follows:
\begin{equation}
\mathcal{U}_{\mathrm{int}}(0,T) = \exp\left(-i \sum_{k=0}^\infty \Omega_k\right),
\label{Magnus_expansion}
\end{equation}
where
\begin{equation}
\Omega_1 = \int_0^T \mathcal{H}_{\mathrm{int}}(s) ds,
\end{equation}
\begin{equation}
\Omega_2 = \frac{1}{2} \int_0^T \int_0^{s_1} [\mathcal{H}_{\mathrm{int}}(s_1),\mathcal{H}_{\mathrm{int}}(s_2)] ds_1 ds_2,
\end{equation}
and so forth.  To simplify analysis, we transform the integration variables to dimensionless form as
\begin{equation}
u_j = \frac{2}{T}s_j - 1,
\end{equation}
so that the domain of integration is now $[-1,1]$. As a result, $\Omega_n \propto T^n$, so the expansion in \refeq{Magnus_expansion} is a power series in $T$.

We say that a sequence which satisfies $\Omega_j \simeq 0$ for $j = 1, \ldots, n$ ``decouples to $n^{\mathrm{th}}$-order'' because the first non-vanishing error is of order $T^{n+1} \left\| \mathcal{H}_{\mathrm{int}} \right\|^{n+1}$.  We use approximate equality because the CAFE sequence we will derive is ultimately limited by numerical precision and physically realizable control fields, which may violate strict equality of the ``decoupling constraints.''  However, we defend the introduction of any approximation errors with our simulation results.  This approach to the dynamical decoupling problem is akin to time-dependent perturbation theory, and is useful when the noisy environment is not strongly perturbing the system and hence the Magnus expansion (in the interaction picture defined above) converges rapidly.

The ``zeroth-order'' constraint which must be satisfied is that the control field itself (in absence of any bath effects) is identity.  This could be modified, if desired, to give some single-qubit rotation, but we do not pursue this possibility here. Hence,
\begin{equation}
\beta(T) = 0 \; (\mathrm{mod} \; 2 \pi).
\end{equation}
This additional constraint must be appended because the initial set of constraints was derived in the interaction picture.  For simplicity, we will use $\beta(T) = 0$ strictly in this work.

The first- and second-order decoupling constraints can be derived from \refeq{H_int} and \refeq{Magnus_expansion} by nullifying any terms which are not identity operations on the qubit.  For example, the first-order constraints ($\propto T$) are
\begin{eqnarray}
\int_{-1}^{1} \sin [\beta (u)] du &=0,
\label{DD_sine1}
\\
\int_{-1}^{1} \cos [\beta (u)] du &=0.
\label{DD_cosine1}
\end{eqnarray}
The second-order constraints ($\propto T^2$) can be derived from the Magnus expansion as
\begin{eqnarray}
\int_{-1}^{1} \int_{-1}^{u_{1}} \left\{\sin [\beta (u_{1})] - \sin [\beta (u_{2})] \right\} du_{1} du_{2} &=0,
\label{DD_sine2}
\\
\int_{-1}^{1} \int_{-1}^{u_{1}} \left\{\cos [\beta (u_{1})] - \cos [\beta (u_{2})] \right\} du_{1} du_{2} &=0,
\label{DD_cosine2}
\\
\int_{-1}^{1} \int_{-1}^{u_{1}} \sin \left[\beta (u_{1}) - \beta (u_{2})\right] du_{1} du_{2}&=0.
\label{DD_2nd_order_crossterm}
\end{eqnarray}
These continuous DD constraints are equivalent to those studied elsewhere, for example in Ref.~\cite{Pasini2009}.

\subsection{A Continuous Control Function that Mimics Uhrig Dynamical Decoupling}
\label{smooth_approximation}
Finding arbitrary time-dependent control functions that satisfy the continuous DD constraints, even at first-order, is not trivial.  We begin our search by studying the cumulative pulse area $\beta_{\mathrm{UDD}} (t)$ for the Uhrig dynamical decoupling sequence (UDD) because this sequence is known to decouple by perturbative order using bang-bang pulses~\cite{Yang08}.  By transforming to the domain  $u \in [-1,1]$, the time-dependent rotation in a UDD sequence of $N$ pulses is given by
\begin{eqnarray}
\beta_{\mathrm{UDD}} (u) & = \pi \int_{-1}^{u} \sum_{j = 1}^{N} \delta\left[u_1 + \cos\left(\frac{j \pi}{N+1}\right)\right] du_1 \nonumber \\
                        &=  \pi \sum_{j = 1}^{N} \Theta\left[
                        u + \cos\left(\frac{j\pi}{N+1}\right)\right],
\end{eqnarray}
where $\delta(u)$ is the Dirac delta function and $\Theta(u)$ is the Heaviside step function.

\begin{figure}
  \begin{center}
  \subfloat{\includegraphics[width=0.5\textwidth]{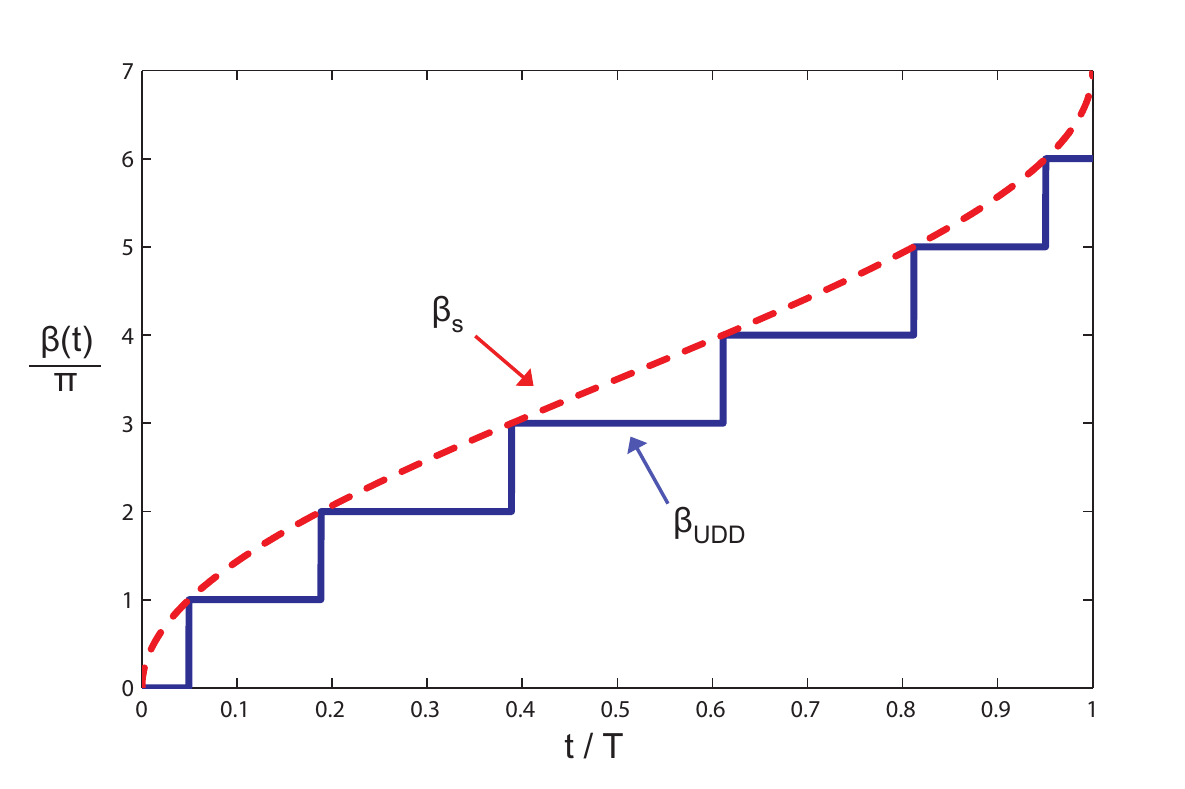}}
  \subfloat{\includegraphics[width=0.5\textwidth]{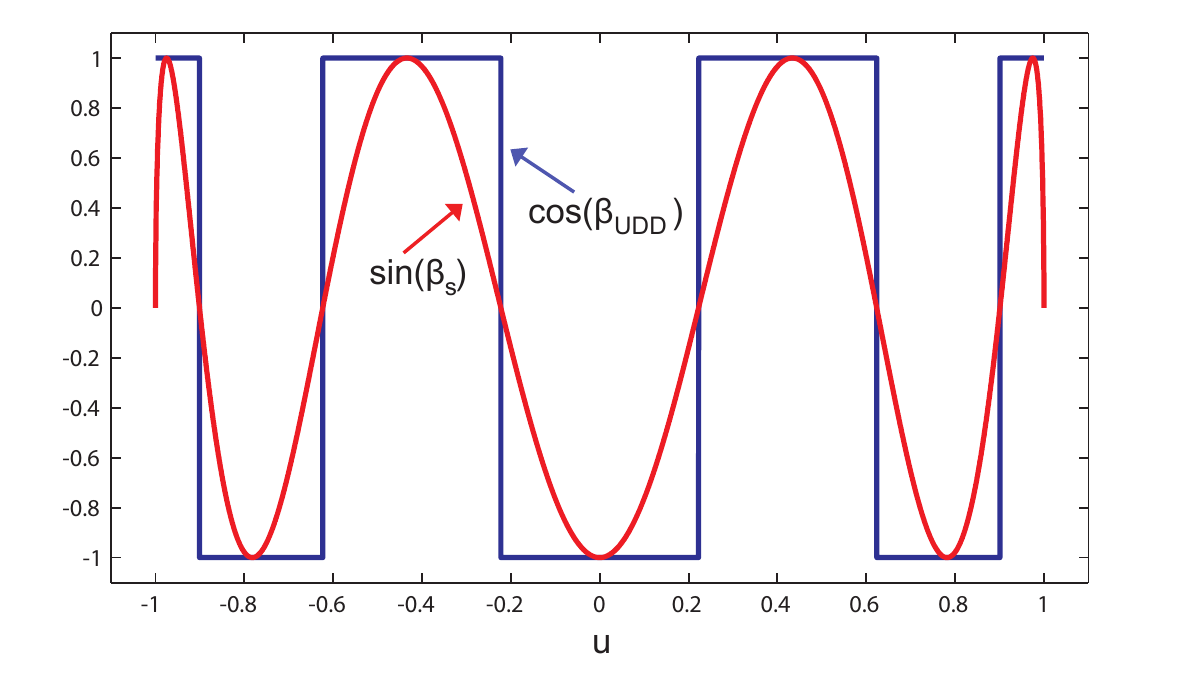}}
  \caption{(a) A smooth approximation to the UDD pulse area over time using a continuous control function. At prescribed instants in time, the UDD sequence applies a $\pi$-pulse, seen here as a jump in the cumulative pulse area. The smooth approximation intersects with the UDD sequence at these pulse times.  The traces are divided by $\pi$ to show the jumps at each $\pi$-pulse as integral steps in this plot.
  (b)
  Comparison between the UDD switching function (blue) and the function $\sin (\beta_{\mathrm{s}} (u)) = \sqrt{1-u^{2}} U_{N}(u)$ for $\beta_{\mathrm{s}}$ (red), where $U_N (u)$ is the $N^{\mathrm{th}}$ Chebyshev polynomial of the second kind.  Because $\sqrt{1-u^{2}}$ is the weight function for this polynomial, it has many of the same orthogonality properties as the UDD switching function.
  }
  \label{CCUDD}
  \label{switching_function}
  \end{center}
\end{figure}

The integrated control function for UDD, $\beta_{\mathrm{UDD}} (u)$, can be well approximated by a smooth function of the form
\begin{equation}
\beta_{\mathrm{s}} (u) = (N+1) \cos ^{-1} (-u),
\label{beta_smooth}
\end{equation}
where $N$ corresponds to the number of pulses in the UDD sequence, which is plotted in \reffig{CCUDD} for $N=6$.  Note that Ref.~\cite{Yang08} makes a convenient variable substitution, which is to operate in a transformed coordinate basis given by $\theta = \cos ^{-1} (-u)$.  In this basis, $\beta_s$ is proportional to $\theta$.  We utilize this transformation later.  By differentiation of \refeq{beta_smooth}, one can determine that the corresponding control function is given by
\begin{equation}
\alpha_{\mathrm{s}} (u) = \frac{N+1}{2 \sqrt{1 - u^{2}}}.
\end{equation}
Moreover, the interaction picture Hamiltonian can now be expressed in terms of Chebyshev polynomials when using this continuous control function since
\begin{equation}
\cos [(N+1) \cos ^{-1} (-u)] = T_{N+1} (u),
\end{equation}
\begin{equation}
\sin [(N+1) \cos ^{-1} (-u)] = \sqrt{1-u^{2}} U_{N} (u),
\end{equation}
where $T_{N+1}$ is the $(N+1)^{\mathrm{th}}$-order Chebyshev polynomial of the first kind and $U_{N}$ is the $N^{\mathrm{th}}$-order Chebyshev polynomial of the second kind.  The latter is particularly noteworthy because $\sqrt{1-u^{2}}$ is the weight function for $U_{N}$, so that $\sqrt{1-u^{2}} U_{N} (u)$ is orthogonal to any polynomial in $u$ of order $N-1$ or lower on the interval $[-1,1]$.  This property allows us to assert that the dynamical decoupling constraints containing just a ``sine'' function, as in \refeq{DD_sine1} and \refeq{DD_sine2}, are made zero by this choice of function~\cite{Yang08}.  We can summarily state this as
\begin{equation}
\underbrace{\int_{-1}^{1} du_{1} \ldots \int_{-1}^{u_{m-1}} du_{m}}_{\times m} \sin \left[\beta_{\mathrm{s}} (u_{k})\right] = 0
\end{equation}
for $m \le N-1$ and $1 \le k \le m$~\cite{Yang08}. Furthermore, half of the corresponding ``cosine'' constraints in the $u$-domain are zero by parity rules, since $T_{N+1}(u)$ has definite parity.  The remaining constraints are not zero, so this control function has a nonzero constraint at first- or second-order.   Figure~\ref{switching_function} compares $\sin \left[\beta_{\mathrm{s}} (u)\right]$ to $\cos \left[\beta_{\mathrm{UDD}} (u)\right]$, also known as the UDD ``switching function'' $y(t)$.  This indicates that $\beta_{\mathrm{s}}$ preserves many of the decoupling properties of UDD, as supported by numerical simulations to be discussed later.

Before proceeding to derive new sequences based on smooth approximations of UDD, we note that the connection of UDD's switching function to Chebyshev polynomials allows an alternative interpretation of its decoupling abilities.  
To arrive at a switching function, we treat the time-dependent operator $B(t)$ as a random classical function, following standard procedures surrounding the Born-Markov approximation.  In a particular trajectory, a qubit in state $c_0\ket{0}+c_1\ket{1}$ evolves to $c_0\ket{0}+\exp(i\phi)\ket{1}$ with the accrued phase during the UDD sequence
\be
\label{smoothstart}
\phi = \int_0^T B(t) y(t) dt.
\ee
This integral may be rewritten
\be
\label{thetasum}
\phi = \frac{2(N+1)}{\pi}\sum_{j=0}^{N+1} (-1)^j w_j  \int_0^{\tau_j} B(\tau) d\tau,
\ee
where the $\tau_j$ are the locations of UDD pulses, i.e.,
$\tau_j = T\sin^2({j\pi}/[2(N+1)]),$
or, on the $u$ axis, $u_j =-\cos({j\pi}/(N+1)$.
The weights are given by $w_j ={\pi}/{2(N+1)}$ for $ j=0,{N+1}$ and $w_j={\pi}/{(N+1)}$ otherwise.
These weight functions are those of Gauss-Lobatto-Chebyshev quadrature, which says that for order $M$,
\be
\label{glc}
\sum_{j=0}^{M} w_j f(u_j) = \int_{-1}^1 \frac{f(u)}{\sqrt{1-u^2}}du + R_M,
\ee
where $R_M$ is a remainder term which vanishes quickly with $M$ and $u_j=-\cos({j\pi}/{M}).$ Hence we associate $M$, the order of Gauss-Lobatto-Chebyshev quadrature, with $N+1$, where $N$ is the number of $\pi$ pulses in a UDD sequence.  The other term we must consider is $(-1)^j$.  This may be written as
$(-1)^j = T_{N+1}(u_j)$,
evident from the formula $T_n(u)=\cos(n\cos^{-1}(u))$.  Therefore, if we define
\be
f(u) = \frac{(N+1)T}{\pi}T_{N+1}(u)\int_{-1}^{u}B[T(v+1)/2]dv,
\ee
then \refeq{thetasum} is exactly in the same form as the left-hand-side of \refeq{glc}.  We may therefore replace our sum with an integral with only higher-order corrections. Then, integration by parts and a return to the $t$ domain yields
\be
\phi = \frac{4}{\pi}\int_{0}^T dt B(t)U_N\biggl(\frac{2t}{T}-1\biggr)\sqrt{\frac{t}{T}\biggl(1-\frac{t}{T}\biggr)} + R_{N+1}.
\label{finalsmoothUDD}
\ee
Recalling that $\sqrt{(t/T)(1-t/T)}=\sqrt{1-u^2}$ is the weight function for the orthogonal polynomials $U_N(-u)$, we see that UDD effectively extracts the $N$th term of an expansion of $B(t)$ in Chebyshev polynomials of the second kind.  Lower-order components of this expansion are decoupled.  Higher-order components still appear in the remainder term $R_{N+1}$.

This result provides two major conclusions.  First, it provides an alternative viewpoint as to how UDD achieves decoupling: UDD eliminates lower order Chebyshev polynomial components of the bath field.  Second, direct comparison of \refeq{finalsmoothUDD} and \refeq{smoothstart} shows that, up to a remainder $R_{N+1}$ which vanishes for high $N$, the discontinuous switching function $y(t)$ is well approximated by the continuous function $(4/\pi)U_N(2t/T-1)\sqrt{(t/T)(1-t/T)}=(4/\pi)\sin[\beta_{\mathrm{s}}(t)],$ as shown in \reffig{switching_function}b.  This connection motivates our effort to find smooth versions of the UDD sequence.

\subsection{Variational Parameters for Improved continuous DD Control Functions}
We will now develop a method for improving the dynamical decoupling properties of the continuous-control sequence given by $\beta_{\mathrm{s}}$.  We consider a technique for expanding the difference $\beta_{\mathrm{s}}-\beta_{\mathrm{UDD}}$ as a Fourier series, where we subsequently truncate the series and tune the remaining terms for optimal performance.  Recent work by Uhrig has adapted the UDD sequence to replace instantaneous pulses with time-dependent control functions, while preserving many of the error-suppression properties~\cite{Uhrig2010}. This followed a mathematical structure derived by Yang and Liu~\cite{Yang08} used to prove that the original UDD sequence works as intended.  Recall the variable substitution $\theta = \cos ^{-1} (-u)$.  In the $\theta$-domain, $\beta_{\mathrm{UDD}}(\theta)$ is a ``staircase'' function while $\beta_{\mathrm{s}} (\theta) = N\theta$ is a linear approximation to $\beta_{\mathrm{UDD}}$.  The difference between the two sequences is a periodic sawtooth wave, which has a Fourier series given by
\begin{equation}
\beta_{\mathrm{s}}(\theta) - \beta_{\mathrm{UDD}}(\theta) = \sum_{j = 1}^{\infty} \frac{\sin (2Nj \theta)}{j}.
\end{equation}

\begin{figure}
  \begin{center}
  \includegraphics[width=\textwidth]{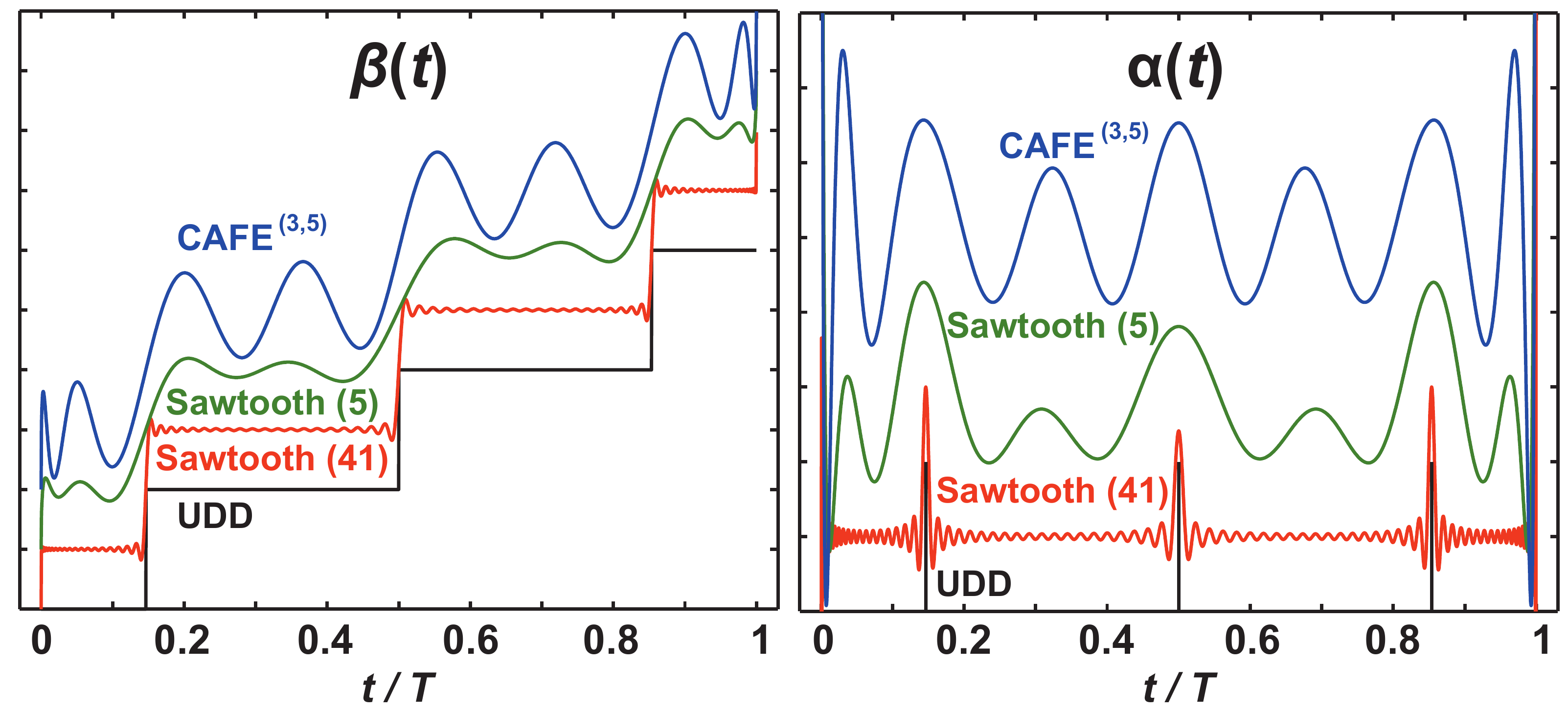}
  \caption{Construction of the CAFE sequence.  The left panel shows $\beta_{\mathrm{UDD}}(t)$ in black, a staircase function with steps of $\pi$.  Using \refeq{triangle}, the red curve closely approximates $\beta_{\mathrm{UDD}}(t)$ as a sum of $\beta_{\mathrm{s}}(t)$ and a 41-term Fourier series for a sawtooth wave (only 20 terms of which are nonzero).  The green curve truncates this series to 5 terms, only two of which are nonzero; an offset is added to the figure for clarity.  The blue curve is the numerical solution for the 5-parameter \rawCAFE{3,5}\ sequence, found using the green curve as a starting guess, again with added offset.  The right panel shows the corresponding $\alpha(t)$ functions for these approximations; the UDD sequence is a series of $\delta$-functions shown as black lines.  The continuous approximations, again offset for clarity, have poles at the endpoints.
  }
  \label{CCUDD_Fourier}
  \label{CAFE_3_5_alpha}
  \end{center}
\end{figure}

The first attempt at improving the control function would be to truncate the Fourier series with the first several terms (as shown in \reffig{CCUDD_Fourier}), but this results in only a modest improvement in decoupling.  However, there is an insight to be gained here, because it turns out that any ``sine'' Fourier components given by $\lambda_p \sin (Np \theta)$ do not affect the ``sine'' DD constraints, in the sense that
\begin{equation}
\label{triangle}
\underbrace{\int_{0}^{\pi} d\theta_{1} \ldots \int_{0}^{\theta_{m-1}} d\theta_{m}}_{\times m} \sin \biggl[\beta_{\mathrm{s}} (\theta_{k}) + \sum_{p} \lambda_{p}\sin (Np \theta_{k})\biggr] = 0
\end{equation}
for $1 \le k \le m$ and any real coefficients $\lambda_{p}$ (see \ref{cancellation_proof}).  We can exploit this property to nullify some of the other DD constraints by varying the $\lambda_p$ parameters without violating the above ``sine'' DD constraints.  In particular, the ``cosine'' DD constraints do depend on the $\lambda_p$ parameters, so we set up a system of nonlinear equations involving the first $m$ constraints that are not already zero by identity and the first $m$ variational parameters ($\lambda_p$).  These equations involve nested integrals over complicated functions, and so finding genuine roots to the equations can be numerically intensive.

The optimized set of parameters we examine more closely is given by five DD constraints that are not identically zero for the control function
\begin{equation}
\beta_{\lambda}^{(5)} (\theta) = 4\theta + \sum_{k=1}^{5} \lambda_{k}\sin (4k \theta),
\label{beta_CAFE}
\end{equation}
where $\lambda_{1} \ldots \lambda_{5}$ are variational parameters we can change to satisfy the five DD constraints.  As discussed previously, this functional form automatically nullifies ``sine'' DD constraints as in \refeq{DD_sine1} and \refeq{DD_sine2}.  Furthermore, this form has definite parity (even), so half of the ``cosine'' constraints can be eliminated by parity rules.  The system of equations we solve for is given by:
\begin{eqnarray}
\label{Lambda5_system_start}
\int_{-1}^{1} \cos [\beta (u_{1})] du_{1} = 0 \\
\int_{-1}^{1} du_{1} \int_{-1}^{u_{1}} du_{2} \sin [\beta (u_{1}) - \beta (u_{2})] = 0 \\
\int_{-1}^{1} \left(u_{1}\right)^2 \cos [\beta (u_{1})] du_{1} = 0 \\
\int_{-1}^{1} du_{1} \int_{-1}^{u_{1}} du_{2} \left(u_{2}\right)^{2} \sin [\beta (u_{1}) - \beta (u_{2})] = 0 \\
\int_{-1}^{1} \left(u_{1}\right)^4 \cos [\beta (u_{1})] du_{1} = 0.
\label{Lambda5_system_end}
\end{eqnarray}

Importantly, the parity rules noted in Section~\ref{smooth_approximation} only apply to equations in the $u$-domain, not in real time.  Still, this set of
equations nullifies all of the $1^{\mathrm{st}}$- and $2^{\mathrm{nd}}$-order constraints in (\ref{DD_sine1}--\ref{DD_2nd_order_crossterm}).  Moreover, several error terms at $3^{\mathrm{rd}}$-, $4^{\mathrm{th}}$- and $5^{\mathrm{th}}$-order are also canceled.

\begin{table}
    \centering
    \begin{tabular}{|c|c|c|}
      \hline
       & Fourier & Optimized Search \\ \hline
      $\lambda_{1}$ & 0 & 0.0017 \\ \hline
      $\lambda_{2}$ & 1 & 0.9121 \\ \hline
      $\lambda_{3}$ & 0 & -0.2869 \\ \hline
      $\lambda_{4}$ & \sfrac{1}{2} & 1.3520 \\ \hline
      $\lambda_{5}$ & 0 & 0.4920 \\ \hline

    \end{tabular}
    \caption{The first five $\lambda_p$ parameters for a truncated Fourier series and the optimized sequence found by numerical solution of five decoupling
    constraints.}
    \label{Lambda_params_table}
\end{table}

The $\lambda_p$ parameters were treated as variables to solve the system of equations in (\ref{Lambda5_system_start}--\ref{Lambda5_system_end}).  The initial approximate solution was taken to be the truncated Fourier series, and Table~\ref{Lambda_params_table} lists the initial parameters and their values after optimization.  Figure~\ref{CAFE_3_5_alpha} shows the optimized continuous control function with five~$\lambda_p$ parameters.  We call such decoupling sequences with optimized parameters CAFE.  To specify precisely which CAFE sequence is being considered, we denote each as \rawCAFE{N,m} where $N$ is the order of the continuous approximation to UDD, as appears in \refeq{beta_smooth} and subsequent references to $\beta_{\mathrm{s}}$ like \refeq{beta_CAFE}.  The quantity $m$ is the number of variational $\lambda_p$ parameters.  For example, \reffig{CAFE_3_5_alpha} shows the control function $\alpha^{\mathrm{(5)}}(t)$ for \rawCAFE{3,5}.

\subsection{Piecing Together Sequences: Splice and Invert}
\begin{figure}
  \begin{center}
  \subfloat{\label{CAFE_chopped}\includegraphics[width=6.8cm]{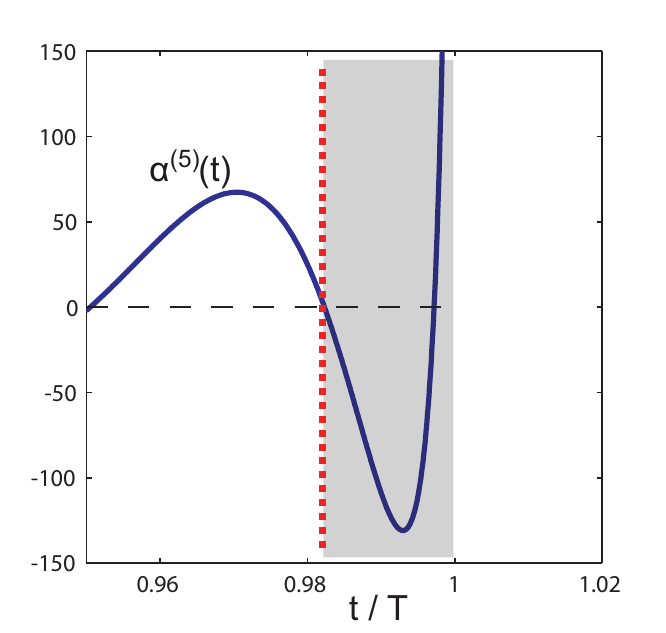}}
  \subfloat{\label{CAFE_x2}\includegraphics[width=6.8cm]{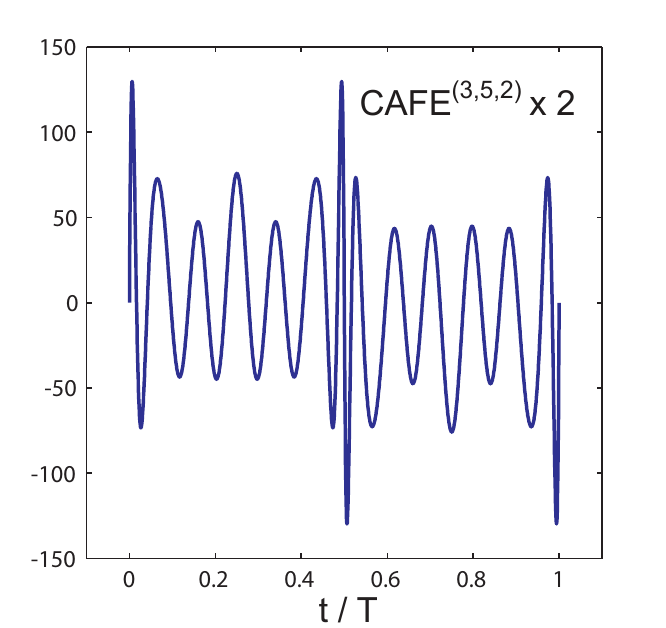}}
  \caption{The \rawCAFE{3,5} sequence can be made finite by removing the beginning and the end of the sequence. (a) The
    \rawCAFE{3,5} sequence ``chopped'' at the second zero from the end of the sequence.  (b) Two iterations of the \rawCAFE{3,5,2} base sequence, showing that the original was chopped at the second and second-to-last zeros.  The second time window is the negative of the first.  This function is finite and smooth at all points.}
  \end{center}
\end{figure}

The previous section described how to improve sequences by inserting $m$ variational parameters and searching for a set that minimizes $m$ previously nonzero DD constraints.  The resulting control function $\alpha^{(m)}(t)$ has the undesirable property of having poles at the start and endpoints, which are impractical to implement in an experiment.  However, the function is very steep here as well, so it seems reasonable to ``chop off'' this section of the sequence with minimal impact on the decoupling properties.  This process is illustrated in \reffig{CAFE_chopped}.  The problem here is that one must also ensure that the net action of the control function is the identity gate.  Otherwise, the sequence will decouple the qubit well, but the state of the qubit will be rotated deterministically around the Bloch sphere.  This implies that the pulse area in $\beta(t)$ that is removed must be equal to $2\pi$.  The pulse area can be calculated readily, but ensuring this area is precisely $2\pi$ in a real experiment seems unreasonable, given how large the first derivative of $\alpha (t)$ is at the endpoints of the sequence.

A rather simple and effective solution is to chop at zeros of $\alpha(t)$ near the start/end of the sequence, as shown in \reffig{CAFE_chopped}.  For $\alpha(t)$ being an even function of time, the zeros lie symmetrically about the sequence, so we can splice together a window of time with control $\alpha(t)$ followed by an equal window of time with control $-\alpha(t)$ which has been chopped in the same manner; for $\alpha(t)$ being an odd function of time, one must time-reverse control in one of the windows.
Figure~\ref{CAFE_x2} shows how to use this method to make a simple bounded sequence by splicing and inverting two segments of the CAFE sequence from \reffig{CAFE_3_5_alpha}.
The resulting sequence has several advantages: (1) it is finite and continuous, and the first derivative is continuous, at all points; (2) the total pulse area is automatically zero since the second half of the sequence reverses any action in the first half; (3) this sequence preserves much of its dynamical decoupling properties and performs quite well in simulation. We expand the notation for this decoupling function to \genCAFE{N,m,r}{L} where $r$ is the root-number at which the sequence is truncated (counting from the end) and $L$ is the number of times the CAFE base sequence is repeated.  For example, \reffig{CAFE_x2} shows a \CAFE{2}{2} sequence.

\section{Comparing different sequences}
    \label{comparison}
We evaluate the performance of the CAFE sequences in comparison to a very similar set of sequences, namely the RUDD sequences of Uhrig and Pasini \cite{Pasini2009,Uhrig2010}.  However, we make two alterations to the original RUDD sequence to make the comparison as appropriate as possible.  First, the RUDD sequence as originally derived \cite{Pasini2009} begins and ends with very sharp pulses of area $2\pi$.  Our numeric studies as well as those of Uhrig and Pasini \cite{Uhrig2010} have shown that these pulses have very little effect on the decoupling abilities of the RUDD sequence.  We therefore omit these pulses in all comparisons shown in this paper.  Another modification we make is to invert the sign of each subsequent pulse in the RUDD sequence.  We denote a RUDD sequence modified in this way by \CRUDD{N}{d}, where $N$ is the number of pulses and $d$ is the duty cycle.

The inversion of subsequent copies of CAFE was introduced to assure continuity, but this process also introduced a further advantage of adding robustness to low-frequency error in the amplitude of the control field.  Such errors will be discussed in more detail in Sec.~\ref{simulation}.  In the case of RUDD with perfect pulses, the modification of inverting each pulse has no effect on the performance of the sequence.  However, it introduces the same robustness to low-frequency control noise, and so therefore provides a closer comparison to the CAFE sequences.

\begin{figure}
    \includegraphics[width=\textwidth]{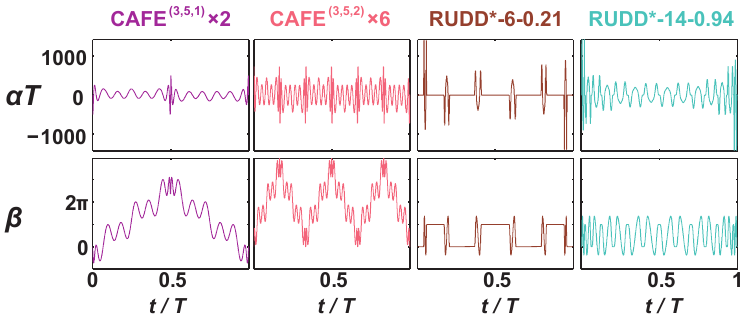}
    \caption{Four sequences with comparable maximum slew rate $\max_t T^2|d\alpha/dt| \approx 7\times 10^5$.  The top row shows the continuous control field $\alpha(t)$ in units of $T$.  The bottom row shows $\beta(t)$, the integral of $\alpha(t)$; this function would look like a square wave from $0$ to $\pi$ for an ideal $\pi$-pulse sequence such as UDD.}
    \label{lowslewseqfig}
\end{figure}

\begin{figure}
    \includegraphics[width=\textwidth]{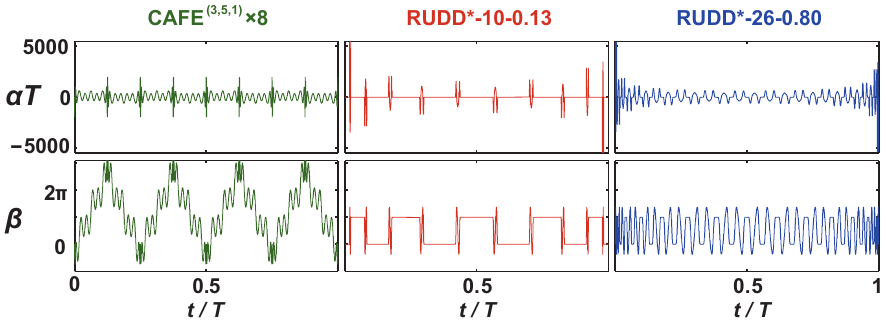}
    \caption{Three sequences with comparable maximum slew rate $\max_t T^2|d\alpha/dt| \approx 10^7$; see \reffig{lowslewseqfig} caption.}
    \label{highslewseqfig}
\end{figure}

In this paper, we will limit our discussion to \CAFE{r}{L}\ sequences, varying only the root number $r$ and the number of repetitions $L$.  We will compare these sequences to \CRUDD{N}{d}\ sequences of different pulse numbers $N$ and duty cycles $d$.  In general, increasing $L$ or reducing $r$ improves the decoupling ability of CAFE, while increasing $N$ or reducing $d$ improves the decoupling ability of \rawCRUDD.  Any of these actions, however, increases the maximum slew rate, $\max_t |d\alpha/dt|$, of the sequence.   For \rawCRUDD\ a high duty cycle $d$ means that pulses are stretched to have longer duration, which reduces slew rate since integrated pulse area is constant ($\pi$-pulses). Limits on slew rate are likely to be the primary reason for using continuous sequences such as CAFE and \rawCRUDD, and therefore we use this criterion to choose which \CAFE{r}{L}\ sequence to compare to which \CRUDD{N}{d}\ sequence.  In particular, for numeric studies we will look at two groups of sequences, each of roughly equal maximum slew rate. One group has a lower slew rate and is summarized in \reffig{lowslewseqfig}; a second with higher slew rate is summarized in \reffig{highslewseqfig}.

\section{Decoupling a qubit from a quantum bath}
    \label{quantum_bath}
\begin{figure}
  \centering
  \includegraphics[width=\textwidth]{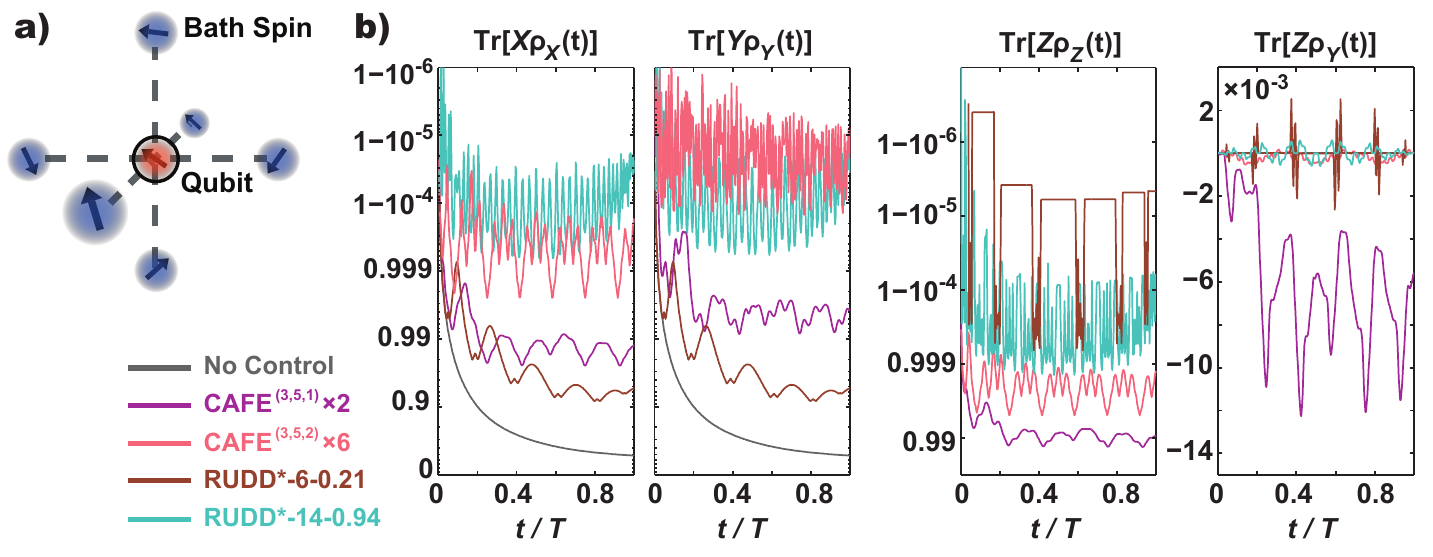}
  \caption{Simulation of decoupling a qubit from spin lattice.
  (a)~Diagram of the spin lattice.  The qubit is surrounded by six bath spins in a cubic lattice configuration.
  (b)~Qubit Bloch vector components as a function of time during decoupling sequences, in the interaction picture, for varying initial conditions.  The four sequences plotted are those in \reffig{lowslewseqfig}.  The grey curves correspond to no control.  For this figure, $\sigma^x$ is abbreviated $X$, etc. Note that in the case of no control, a qubit beginning in state $\rho_z(0)$ will not change, hence the lack of grey curve on the logarithmic plot of $\mathrm{Tr}[Z\rho_Z(t)].$
  This particular simulation used $C_j=1/T$ for all bath spins $j$.
  }
  \label{Quantum_bath}
\end{figure}

Our first evaluation of the CAFE sequence considers a fully quantum mechanical description of a bath.  For this we employ the canonical central-spin problem, in which our qubit is coupled to 6 spins which are themselves coupled via dipole-dipole interactions, i.e.
\begin{eqnarray}
B_0 &= 4J \sum_{j \ne k} \frac{r_{jk}^2\vec{I}_j\cdot\vec{I}_k-3(\vec{r}_{jk}\cdot \vec{I}_j)(\vec{r}_{jk}\cdot\vec{I}_k)}{r_{jk}^5}\\
B_Z &= \sum_j C_j I^z_j,
\end{eqnarray}
where $\vec{I}_j$ is the spin vector (using $I=1/2$) for the bath spins and $\vec{r}_{jk}$ is the spatial vector connecting them.  The cubic geometry we choose is shown in \reffig{Quantum_bath}a.  This type of simulation tests the DD sequences against a system with more degrees of freedom [c.f.~(\ref{DD_sine1}--\ref{DD_2nd_order_crossterm})] than the Born-Markov classical bath simulations we present later.  However, these additional degrees of freedom also complicate analysis, so we present no analytic expectation for the result.   We demonstrate only that CAFE successfully decouples a qubit from a fully-quantum environment in a manner highly comparable to \rawCRUDD.

The plot in \reffig{Quantum_bath}b shows the projection of the qubit Bloch vector in the interaction picture after averaging over the bath spins.  Specifically, this simulation initializes into state
\begin{equation}
\rho_j(0)=\frac{\sigma^I+\sigma^j}{2}\otimes \frac{\id}{2^6},
\end{equation}
where $\sigma^j$ is either a Pauli matrix or single-qubit identity ($j=I,x,y,z$) and $\id$ is the bath identity matrix. Numeric integration then solves $d\rho_j/dt=-i[H,\rho_j]$ for each $j=x,y,z$, and the three qubit directions $k=x,y,z$ are plotted as $\mathrm{Tr}[\sigma^k \rho_j(t)]$ each time $t$ for varying initial conditions $j$.

To evaluate infidelity, we consider the procedure of quantum process tomography (QPT), in which the final state of the qubit, after the sequence is complete at time $T$, may be written
\begin{equation}
\rho(T) = \sum_{j,k=\{I,x,y,z\}} \chi_{jk}\sigma^j \rho(0) \sigma^k.
\end{equation}
 In single-qubit QPT, measurements of the three components of the Bloch with four initial conditions---$\rho_x$, $\rho_y$, $\rho_z$, and $\rho_{-z}$---are sufficient to construct the matrix $\chi_{jk}$ in general~\cite{ncbook}, but since this system features unitary evolution prior to the final trace, projections for initial condition $\rho_{-z}$ are exactly the negative of projections for initial condition $\rho_{z}$.  Further, for our particular interaction and bath, the Ising-like symmetry assures $\Tr[\sigma^y\rho_x(t)]=\Tr[\sigma^z\rho_x(t)]=0$ and $\Tr[\sigma^x\rho_j(t)]=0$ for all $j\ne x$.  Hence the four choices of projection and initial condition shown in \reffig{Quantum_bath}b are sufficient to construct $\chi_{jk}$.  The infidelity for dynamical decoupling is defined as $1-\chi_{II}$; which may be interpreted as the probability that something other than identity happened to the qubit.  In general, for a single qubit in which the process is unitary evolution involving a bath followed by tracing that bath, one may readily show
 \begin{equation}
 \infid = 1-\chi_{II} = \frac{1}{4}\biggl\{3-\sum_{j=\{x,y,z\}}\Tr[\sigma^j\rho_j(T)]\biggr\}.
 \end{equation}

\begin{figure}
  \centering
  \includegraphics[width=0.8\textwidth]{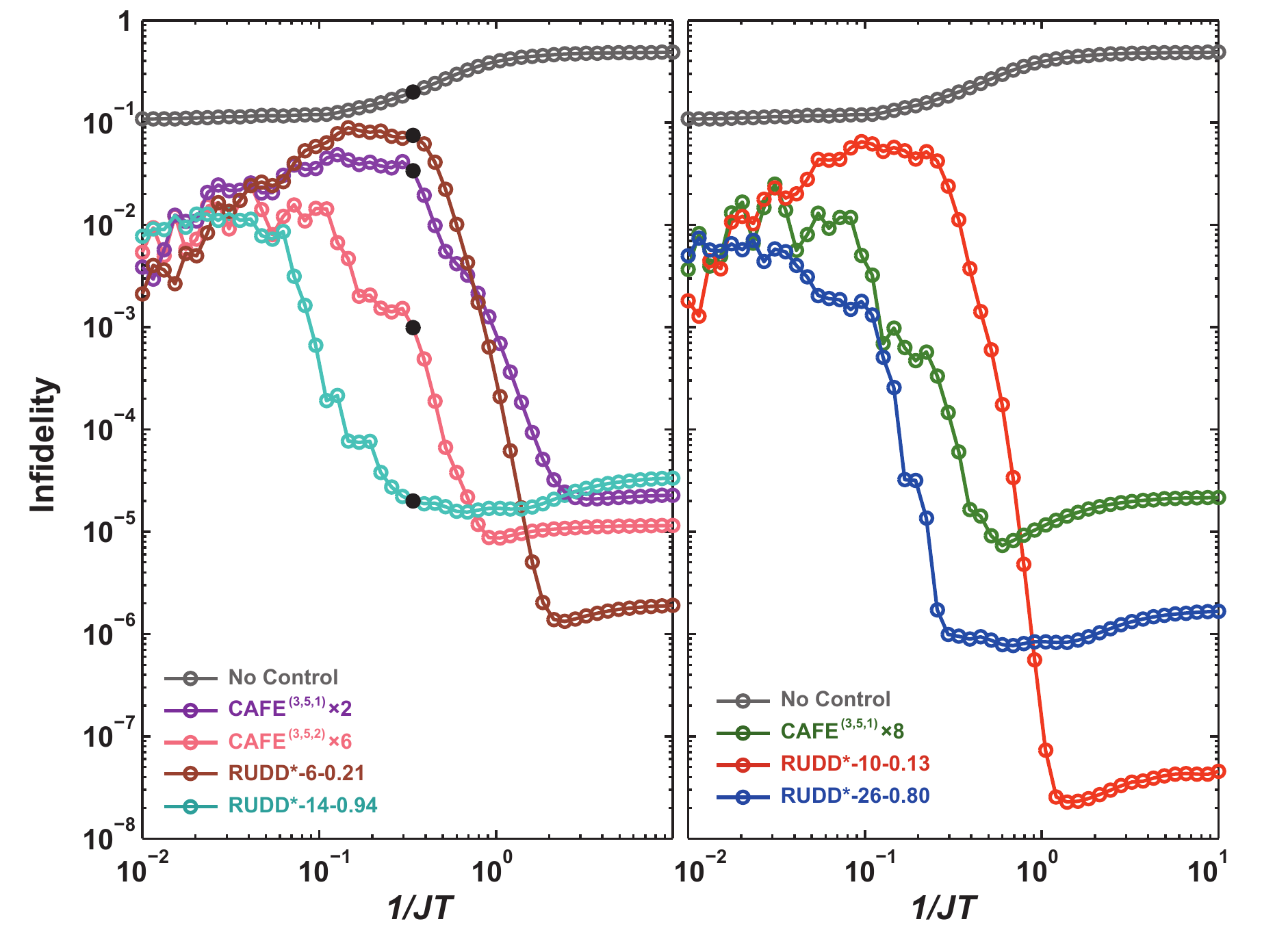}
  \caption{Infidelity when decoupling from quantum bath with various sequences.  The process infidelity is plotted as the overall timescale of the bath dipole-dipole coupling, $1/J$, is varied.
  (Left) Low slew-rate sequences shown in \reffig{lowslewseqfig}.
  (Right)~High slew-rate sequences shown in
  \reffig{highslewseqfig}.  The points filled in with black are those corresponding to the trajectories of \reffig{Quantum_bath}.}
  \label{qbath_sweep}
\end{figure}

 Figure~\ref{qbath_sweep} shows the infidelity as a function of $J$, the overall strength of the dipole-dipole coupling within the bath, for both the low-slew-rate sequences and high-slew-rate CAFE and \rawCRUDD\ sequences shown in figures~\ref{lowslewseqfig} and \ref{highslewseqfig}.  Varying $J$ results in varying the correlation time of the bath.  At low $J$, the bath is very slow, and here dephasing without DD is maximal and determined entirely by the number of possible configurations of the bath spins.  As $J$ increases, even without DD the infidelity decreases due to motional narrowing, in which spin-flips within the bath reduce its coupling to the qubit.  With CAFE and \rawCRUDD\ sequences, it is clearly seen that DD strongly reduces the infidelity at low $J$, corresponding to long correlation times.  The efficacy of decoupling reduces, at varying rates depending on the sequence, as $J$ is increased; improved fidelity at even higher $J$ is again due to motional narrowing effects.

 Figure~\ref{Quantum_bath} shows that \rawCRUDD\ sequences and CAFE sequences are qualitatively different in how they decouple in time; in particular RUDD remains pulsed, while CAFE is fully continuous.  However, \reffig{qbath_sweep} shows that they have roughly the same decoupling power per slew rate as $JT$ is varied.  To analyze their response to varying bath correlation times more quantitatively, we shift to a classical, Born-Markov type bath and a filter function analysis.

\section{Filter function analysis}
    \label{filter_function}
When studying the effects of noise on information, often it is instructive to analyze the system in the frequency domain.  Dynamical decoupling schemes are effectively high-pass filters, meaning they suppress low-frequency components of the system-environment coupling while not reducing high-frequency noise.  The ``cut-off frequency'' between these regions is often closely related to the rate at which one can apply control pulses, but the precise relationship is less clear in examples like UDD, where pulses are not equidistant.  To accurately describe how a decoupling sequence protects against noise with a given spectral density, we characterize each sequence by a ``filter function''~\cite{Biercuk2011}, which characterizes the degree to which noise at a given frequency is suppressed.  Filter functions are a meaningful way to compare different DD sequences because they provide insight into which method is best suited to a given noise spectrum.

We require that the filter function be a linear response which is independent of the magnitude of the environment noise, but this is only appropriate under certain conditions.  First, we invoke the Born-Markov approximation and represent the dephasing bath with a scalar function $B(t)$, so that the Hamiltonian is now
\begin{equation}
\mathcal{H}(t) = B(t) \sigma^z + \alpha(t) \sigma^x,
\label{BM_Hamiltonian}
\end{equation}
where $B(t)$ is a stationary random noise field with zero mean and spectral density $S(\omega)$ given by
\begin{equation}
S(\omega) = \int_{-\infty}^{\infty} dt \langle B(t) B(\tau) \rangle \cos(\omega (t-\tau)).
\label{noise_density}
\end{equation}
Second, we assume that the magnitude of the bath noise is sufficiently weak that we can calculate the qubit infidelity as a linear function of $S(\omega)$ (see \ref{ff_general}).  Under these conditions, we may then define the filter function $F(z)$ by stating that the qubit infidelity $\infid = 1 - \chi_{II}$, or total error, is given by
\begin{equation}
\infid = \int_0^{\infty} \frac{d\omega}{2 \pi \omega^2} S(\omega) F(\omega T).
\end{equation}

The filter function for a given decoupling sequence can be calculated numerically in simulation, by modeling the noisy environment with a sinusoid.  By measuring the degradation in qubit fidelity averaged over an ensemble of random phases for the sinusoid at frequency $\omega$, one obtains the linear response $F(\omega T)$.   For control fields without a $\sigma^z$-component, the filter function may be theoretically expressed as
\be
\label{fftheory}
F(\omega T) = \left|\omega T\int_0^1 e^{i[\omega Tu+\beta(Tu)]}du \right|^2.
\ee
For a general derivation of this expression in the continuous case, see \ref{ff_general}.  This expression clearly reduces to the canonical expression for the filter function derived for sequences composed of ideal $\pi$-pulses such as UDD~\cite{Uhrig 07,Biercuk2011}.  For such sequences, $e^{i\beta(t)}=\cos[\beta(t)]=y(t)$ is the switching function.  Note that if the smooth approximation for $y(t)$ corresponding to UDD, i.e. $(4/\pi) U_N(2t/T-1)\sqrt{(t/T)(1-t/T)}$ as discussed in Sec.~\ref{smooth_approximation}, is substituted into \refeq{fftheory}, the integral may be analytically performed to yield $F(\omega T)\approx 16(N+1)^2J_{N+1}^2(\omega T/2)$ for UDD-$N$, an expression which yields a very close approximation to the exact UDD filter function, as stated without derivation in Ref.~\cite{Uhrig 07}.

\begin{figure}
  \centering
  \includegraphics[width=0.8\textwidth]{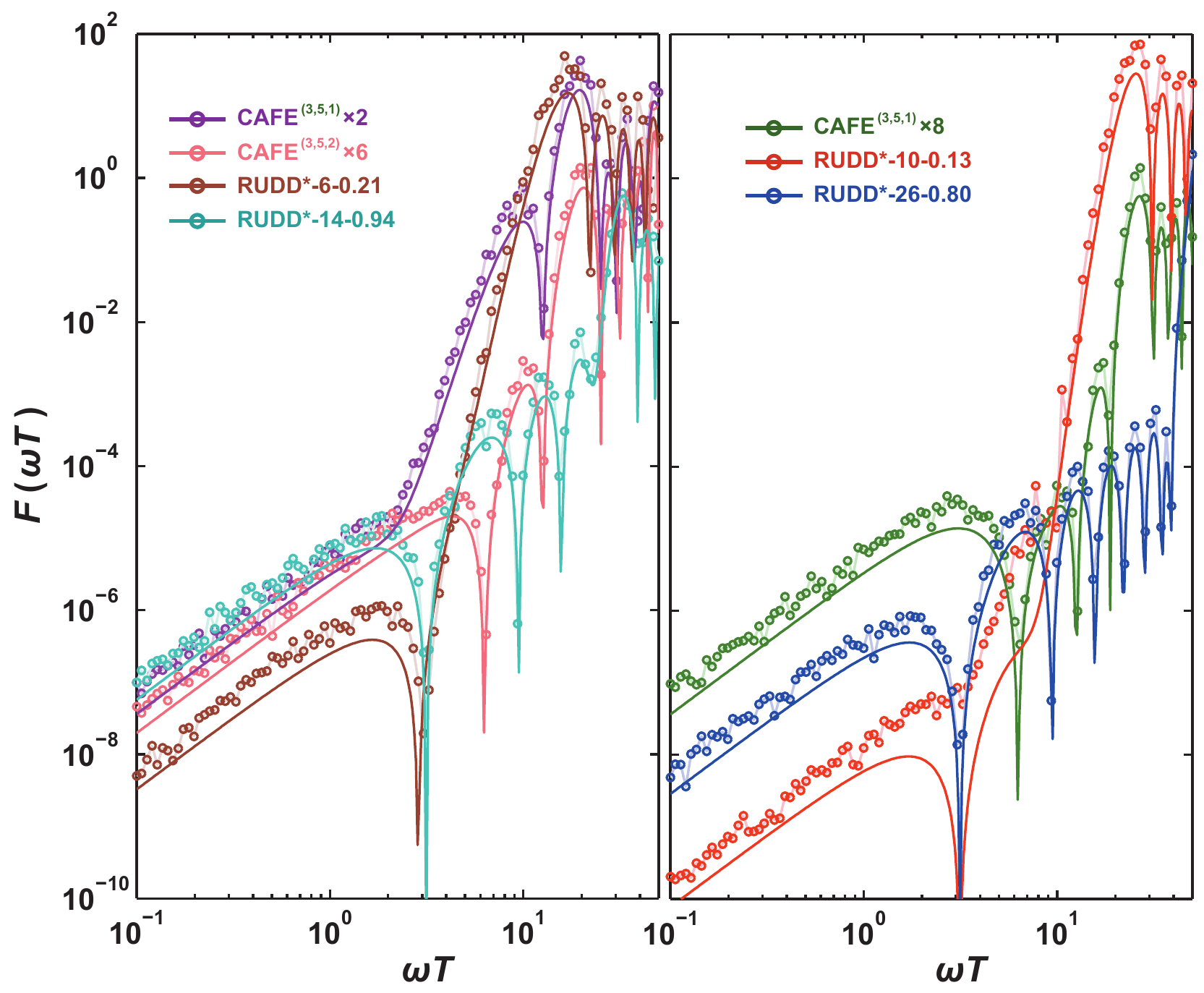}
  \caption{Filter functions calculated from simulation (open circles) and numeric integration (solid lines) following \refeq{fftheory}.  The ``noise'' in the Born-Markov approximation is a sinusoid, so that sweeping frequency over
    many trials of the decoupling sequence reveals the degradation in fidelity, and hence the filter function $F(\omega)$.}
  \label{Filter_function_simulation}
\end{figure}

Figure~\ref{Filter_function_simulation} shows filter functions comparing different versions of CAFE and RUDD, calculated both with explicit integration and via Monte Carlo simulation.
Examining \reffig{Filter_function_simulation}, we see that despite dissimilar time-domain construction, CAFE provides an intermediate limit between low-$N$, low-duty-cycle RUDD sequences and high-$N$, high-duty-cycle RUDD sequences.  The sharp dips in the filter function are quite different between the two cycles, but these play little role in realistic noise bath models which typically integrate out such features.  We turn to such models next.

\section{Decoupling a qubit from Born-Markov classical baths}
    \label{simulation}
To further analyze the effectiveness of CAFE sequences in comparison to \rawCRUDD\, we simulate the total evolution of arbitrary qubit states coupled to classical dephasing baths.  As in the filter function analysis, we invoke the Born-Markov approximation leading to the scalar noise field $B(t)$ in \refeq{BM_Hamiltonian} with noise spectral density $S(\omega)$ given by \refeq{noise_density}.  We now consider bath noise with both Gaussian and Lorentzian spectral density.  The reason we consider a classical bath is that by doing so, we can readily define parameters such as absolute magnitude and correlation time for a scalar noise spectral density, as opposed to the quantum bath which has far too many degrees of freedom to permit simplified analysis.

Our simulations incorporate two sources of noise: environmental dephasing noise and noisy control fields.  Control noise is inevitable in a practical experimental setting.  Even a sophisticated decoupling sequence could yield mediocre results if such control noise is not taken into account; moreover, a DD sequence could possibly introduce more noise than it corrects.  Due to these possibilities, we investigate the effect of control noise on CAFE and \rawCRUDD\ by simulating their decoupling effectiveness with and without noise added to the control functions.

\begin{figure}
    \centering
    \includegraphics[width=\textwidth]{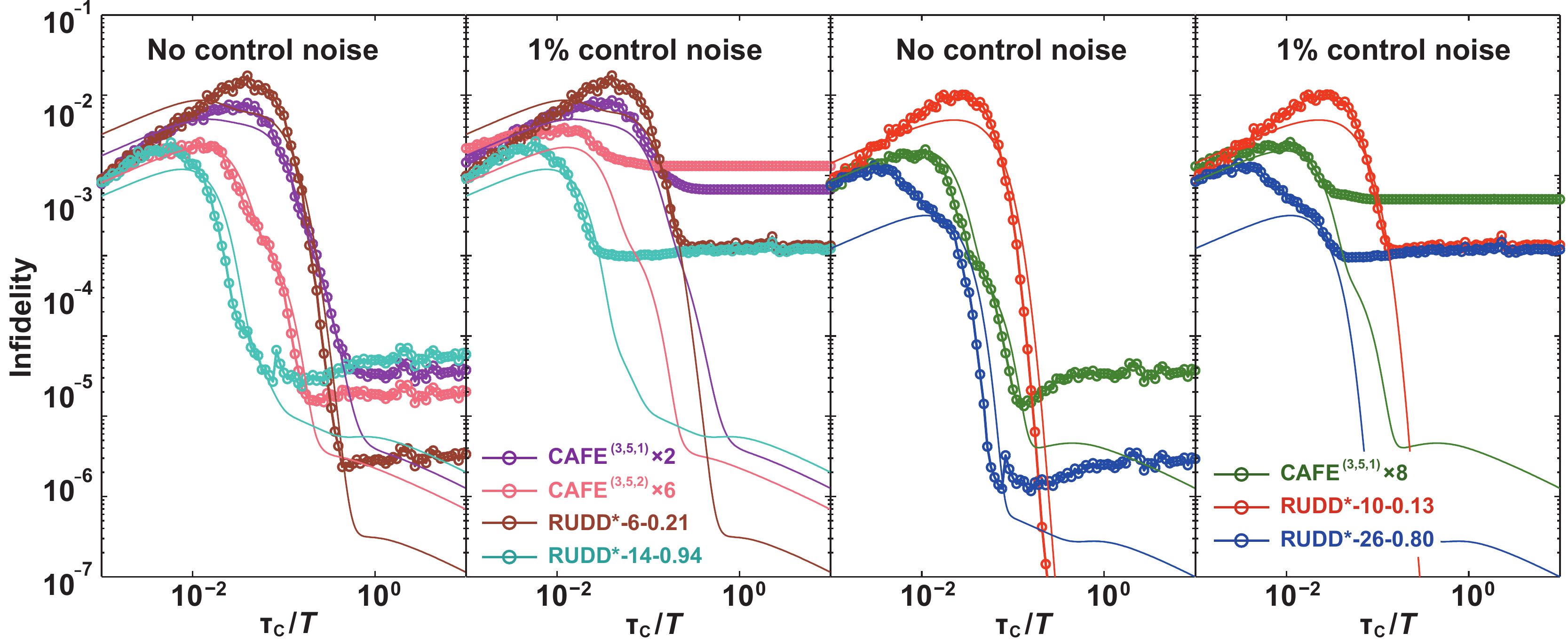}
    \caption{Final state fidelity for decoupling sequences, with environment dephasing noise treated as Born-Markov classical bath with Gaussian $S(\omega)$.  Simulation results are shown with and without noisy controls, for both low slew-rate and high slew-rate sequences.
    The solid lines show numeric integrals of the numerically integrated filter functions (solid lines of \reffig{Filter_function_simulation}) with the spectral density targeted by the simulated dephasing noise.  No modification is made to the solid lines for the plots showing control noise.} \label{Gauss_CAFE}
\end{figure}
\begin{figure}
    \includegraphics[width=\textwidth]{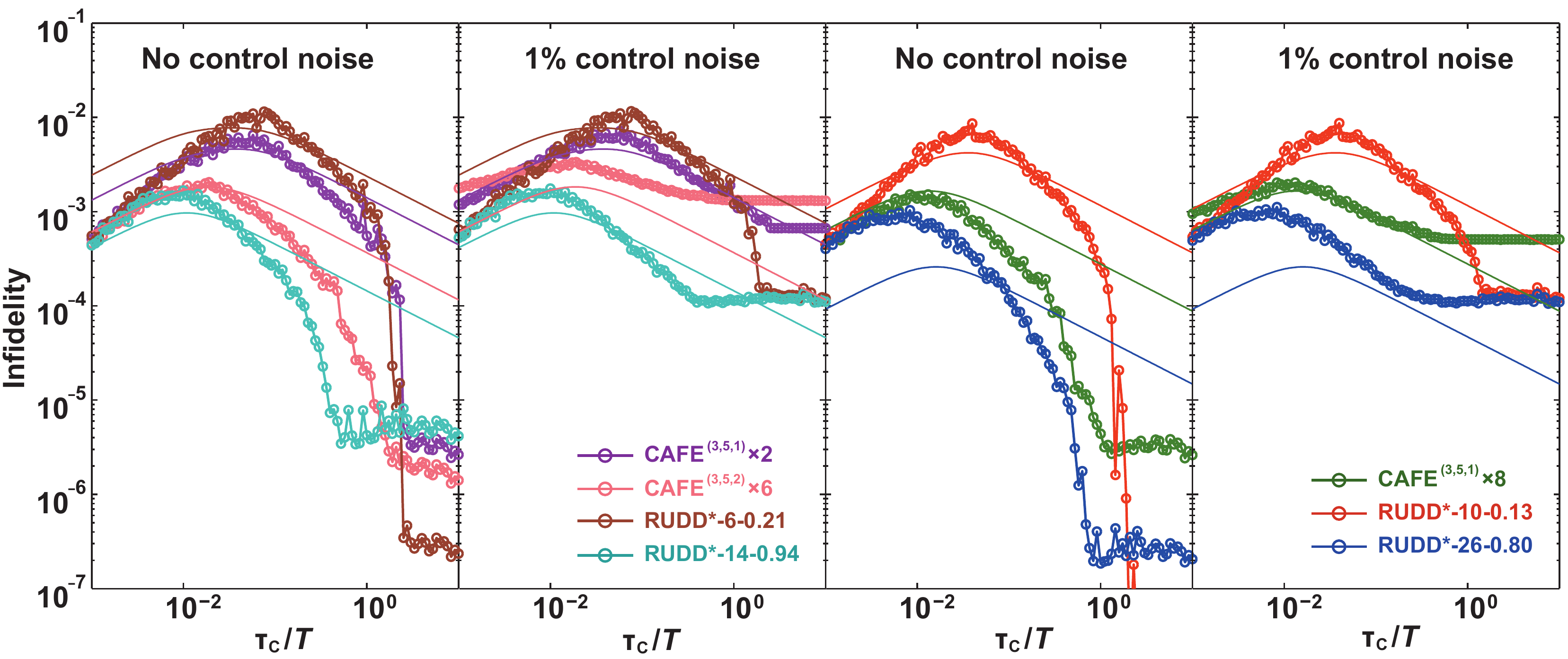}
    \caption{Similar simulation results to \reffig{Gauss_CAFE}, except with environmental dephasing noise treated as Born-Markov classical bath with Lorentzian $S(\omega)$.}
    \label{Lor_CAFE}
\end{figure}
Figure~\ref{Gauss_CAFE} shows the behavior of our example CAFE and \rawCRUDD\ sequences under Born-Markov noise with a spectral density of Gaussian form, i.e. $S(\omega)\propto \exp[-(\omega\ts\tau{c})^2]$, both with perfect control and with 1\% control noise.  Figure~\ref{Lor_CAFE} shows a similar simulation with the spectral density of Lorentzian form, i.e. $S(\omega)\propto 1/[1+(\omega\ts\tau{c})^2]$.  The control noise is modeled as a time-dependent fluctuation on the amplitude of the control field $\alpha(t)$, taken with root-mean-square magnitude 1\% of $\alpha(t)$ and a Gaussian spectral density with correlation time $T/2$ for the plots shown.  Theoretical curves, found by numeric integration, are shown for comparison; the approximations inherent in these curves fail most prominently at very low and very high correlation times $\ts\tau{c}$.   In the case of no control noise, CAFE's performance is roughly intermediate between the \rawCRUDD\ sequences, as already evident from the filter functions analysis of \reffig{Filter_function_simulation}.  One should note the strong similarity between the results of \reffig{Gauss_CAFE} with no control noise and those of \reffig{qbath_sweep}; the nearly identical behavior gives a clear example of a fully quantum bath behaving identically to a classical one, supporting the frequent use of the Born-Markov approximation to describe spin baths.

As control noise is added, we see that all sequences roughly follow their expected behavior at short correlation times, where decoupling is already least effective. In the regime of long correlation times, where DD performs best, a noise-floor is reached which depends on the sequences; this noise-floor is consistently better for the \rawCRUDD\ sequences.  It is tempting to associate this lower noise floor with a lower duty cycle for the \rawCRUDD\ sequences with respect to CAFE's 100\%, but this conclusion is questioned by the highly similar behavior of \rawCRUDD\ sequences of very different duty cycle.  As discussed in Sec.~\ref{comparison}, the noise floor of the RUDD sequences without the modification of alternating the sign of pulses, not shown, is much worse.  This simple method of alternating pulse signs has certainly added some robustness, but many improvements in optimization against control-noise are certainly possible in future work.

\section{Conclusion}
    \label{conclusion}
The CAFE sequence originated from the analysis of UDD as if it were continuous.  Using principles from numerical quadrature, we showed that UDD's essential behavior could be captured by continuous approximations to its normally discontinuous description.  The continuous approximation to UDD provided a heuristic for continuous sequence design.   The RUDD sequences follow a similar path, but are based on a different heuristic~\cite{Pasini2009,Uhrig2010}.  Both sequence families can be extended to higher order.  A promising feature of these types of sequences appears to be that, for these two families at least, sequences can be locally altered to eliminate short-time spiking behavior without appreciable detriment to their decoupling ability.  Although different heuristics were employed, the two sequence families behave remarkably similarly, as we have demonstrated using a variety of analyses.

The reason analysis of continuous-control decoupling is more complicated than the analysis of ``bang-bang'' sequences is because noise can disturb the qubit during control operations.  More than a complication of analysis, however, this situation is a reality of experiment.  We believe, based on our results, that filter functions remain relevant for continuous sequences, and they provide an intuitive picture for what components of the environment noise are suppressed.  Furthermore, Monte Carlo simulations of semi-classical noise bath spectra and fully-quantum spin lattices indicate how well a sequence performs at protecting quantum information.  By adding noise to the control field, we can test another source of error, the experimenter and the experimenter's equipment, and how well the sequence copes.  We hope the heuristics underlying the construction of CAFE and the characterization methods we have used may be useful to future research in continuous dynamical decoupling.

\section*{Acknowledgements}
Sponsored by United States Department of Defense.
The views and conclusions contained in this document are those of the authors and should not be interpreted as representing the official policies, either expressly or implied, of the United States Department of Defense or the U.S. Government.  Approved for public release, distribution unlimited.
\appendix
\section{Proof that Fourier series components do not affect ``sine''-like DD constraints}
    \label{cancellation_proof}
We begin with the definitions from before that
\begin{equation}
\beta_{\mathrm{s}} (t) = (N+1) \cos ^{-1} (-u),
\end{equation}
which in $\theta$-domain ($\theta = \cos ^{-1} (-u)$) becomes
\begin{equation}
\beta_{\mathrm{s}} (\theta) = (N+1) \theta.
\end{equation}
We can now define a control function with $m$ $\lambda$-parameters given as
\begin{equation}
\label{beta_def}
\beta_{\lambda}^{(m)} (\theta) = \beta_{\mathrm{s}} (\theta) + \sum_{k} \lambda_{k}\sin [(N+1)k \theta].
\end{equation}
$\beta_{\lambda}^{(m)}$ is a control sequence with $m$ variational parameters designed to cancel $m$ DD constraints not zero for $\beta_{\mathrm{s}}$ alone.  Here we show that this construction will not affect the ``sine''-like DD constraints given by
\begin{equation}
\underbrace{\int_{0}^{T} dt_{1} \ldots \int_{0}^{t_{m-1}} dt_{m}}_{\times m} \sin [\beta (t_{k})] = 0
\end{equation}
for $1 \le k \le m$. Using integration by parts, the above set of constraints can be reduced to the system of equations
\begin{equation}
\int_{0}^{T} t^{n} \sin (\beta) dt = 0
\end{equation}
for $n = 0, 1, \ldots, (m -1)$.  Working in the $\theta$-domain, this set of constraints is equivalent to
\begin{equation}
\int_{0}^{\pi} [1-\cos (\theta)]^{n} \sin (\beta) \sin (\theta) d\theta = 0,
\end{equation}
which is also equivalent to
\begin{equation}
\int_{0}^{\pi} \cos (p\theta) \sin (\beta) \sin (\theta) d\theta = 0
\end{equation}
for $p = 0, 1, \ldots, (m -1)$. By substituting the expression in \refeq{beta_def},
\begin{eqnarray}
\fl
\int_{0}^{\pi} \cos (p\theta) &\sin [\beta_{\lambda}^{(m)}] \sin (\theta) d\theta =
\\\fl
&\phantom{+}\int_{0}^{\pi} \cos (p\theta) \sin (\theta) \sin [(N+1)\theta] \cos \biggl[\sum_{k} \lambda_{k}\sin ((N+1)k \theta)\biggr]  d\theta \nonumber
\\\fl
&+\int_{0}^{\pi} \cos (p\theta) \sin (\theta) \cos [(N+1)\theta] \sin \biggl[\sum_{k} \lambda_{k}\sin ((N+1)k \theta)\biggr]  d\theta. \nonumber
\end{eqnarray}
Note that
$\cos [\sum_{k} \lambda_{k}\sin ((N+1)k \theta)]
$
can be written as the infinite sum
$
\sum_{j = 0}^{\infty} a_{j} \cos((N+1)j\theta)$
for some real coefficients ${a_{j}}$ since $\cos(\phi)$ is an even function in $\phi$.  Similarly,
$
\sin[\sum_{k} \lambda_{k}\sin ((N+1)k \theta)]
$
can be written as
$
\sum_{j = 1}^{\infty} b_{j} \sin((N+1)j\theta)
$
for some real coefficients ${b_{j}}$ since $\sin(\phi)$ is an odd function in $\phi$.  The DD constraints are therefore satisfied if
\begin{equation}
\int_{0}^{\pi} \cos(p\theta) \sin(\theta) \sin[A(N+1)\theta] \cos[B(N+1)\theta] d\theta = 0
\end{equation}
for integers $A$,$B$ satisfying $A > 0$, $B \ge 0$, and $p < N$. This equation can be directly evaluated to always be true under these conditions.

\section{Derivation of filter function for continuous sequences}
    \label{ff_general}

In this appendix we derive a general expression for the filter function of a continuous DD sequence.  We generalize slightly from the main text, allowing a Hamiltonian of the form
\be
H(t) = 2\epsilon B(t)S^z + \vecalpha(t)\cdot\vec{S}.
\ee
Here $\vecalpha$ is a vector of control fields, one for each component, and $\vec{S}$ is the spin vector, i.e. $S^j = \sigma^j/2$.  The unitless parameter $\epsilon$ is in place to keep track of orders of bath interactions, since the filter function is defined only at the lowest nonvanishing order.  We use $\epsilon$ to \emph{define} that the filter function $F(z)$ in terms of the infidelity $\infid$ at vanishing $\epsilon$.  Again, $\infid$ is defined defined as $1-\chi_{II}$ for the QPT-derived matrix $\chi$, due to evolution at finite $\epsilon$.  Then $F(z)$ is defined so that
\be
\lim_{\epsilon\rightarrow 0} \epsilon^{-2} \infid = \int \frac{d\omega}{2\pi\omega^2} S(\omega) F(\omega T).
\ee
This type of $\epsilon$-based definition is not required for pulses composed of $\delta$-function pulses, but in the continuous case where noise occurs during the control, the limit as $\epsilon\rightarrow 0$ is critical for the linear-response theory upon which the filter function is based.

We begin by transforming to an interaction picture, for which we need the control unitary $U_0(t)=\mathcal{T}\exp[-i\int_0^t \vecalpha(\tau)\cdot\vec{S}]d\tau$.  In terms of this unitary, our interaction picture Hamiltonian is $\tilde{H}(t)=U_0^\dag(t) H(t) U_0^{\phantom{\dag}}(t)$.
If $\vecalpha(t)$ changes in direction over time, $U_0(t)$ in general may describe complicated trajectories over the Bloch sphere.  In this most general case, it is not clear how to arrive at an analytic form for a filter function, and one generally must rely on numeric solvers.  Here, we consider the simple case that $\vecalpha(t) = 2\alpha(t)\dircon$, for constant direction described by unit vector $\dircon$.  The CAFE and RUDD sequences considered in this paper have all used $\dircon$ along the $x$ direction; our definition of infidelity and assumption about the bath interaction indicate that any direction along the Bloch sphere equator is equivalent.  In this appendix, we generalize slightly by allowing a constant $z$-component to $\dircon$ as well.  Then,
\begin{equation}
\fl
\tilde{H}(t) = 2\epsilon B(t) \{S^z\cos[\magcont]+\dircon^z\dircont\cdot\vec{S}\{1-\cos[\magcont]\}
             -\vec{S}\cdot \hat{\vec{z}}\times\dircont\sin[\magcont]\},
\end{equation}
using $\beta(t)$ as defined by \refeq{betadef}.

We consider a single-qubit density matrix in the interaction picture
\be
\rho(t) = \frac{1}{2} + \sum_{j}\rho^j(t) S^j,
\ee
which evolves via $\tilde{H}(t)$, to lowest nonvanishing order in time-dependent perturbation theory with ensemble averaging and standard Born-Markov separation of timescales, as
\be
\frac{d\rho}{dt} = -\int_0^t d\tau \langle [\tilde{H}(t),[\tilde{H}(\tau),\rho(\tau)]]\rangle.
\ee
Correspondingly each Bloch-vector component $\rho^j$ evolves according to the following, where each repeated superscript is summed:
\begin{equation}
\eqalign{
\frac{d\rho^j}{dt} &= -\int_0^t d\tau 2\Tr\{S^j \langle [\tilde{H}(t),[\tilde{H}(\tau),S^k]]\rangle \}\rho^k(\tau)
\\
&= \epsilon^2\int_0^t d\tau \int\frac{d\omega}{2\pi} \cos[\omega(t-\tau)] S(\omega) M^{k\ell}(\tau)M^{\ell j}(t)\rho^k(\tau)
\\
&= -\epsilon^2\int_0^t d\tau \int\frac{d\omega}{2\pi} \cos[\omega(t-\tau)] S(\omega) \Gamma^{jk}(t,\tau)\rho^k(\tau),}
\end{equation}
where
\begin{equation}
\eqalign{
M^{k \ell}(t)
=2\{\varepsilon^{zk\ell}\cos[\magcont]
       +\dircon^z\dircon^m\epsilon^{mk\ell}\{1-\cos[\magcont]\}\\
       \hspace{1in} +[\dircon^k\delta^{z\ell}-\delta^{kz}\dircon^{\ell}]\sin[\magcont]\}.}
\end{equation}
Here  $\varepsilon^{jk\ell}$ is the fully asymmetric Levi-Cevita tensor density and $\delta^{jk}$ is the identity matrix.

Now, we time-integrate $\rho^j$ to lowest order in $\epsilon$, to find
\begin{equation}
\fl\eqalign{
\rho^j(T)
 =\biggl[\id-\epsilon^2\int_0^T dt \int_0^t d\tau \int \frac{d\omega}{2\pi} \cos[\omega(t-\tau)]S(\omega)\Gamma^{jk}(t,\tau)\biggr]\rho^k(0)+\Or(\epsilon^4).}
\end{equation}
We use this expression in QPT, which in general gives
\be
\eqalign{
\chi_{II}(T) = \frac{1}{8}\biggl[&2+2\rho_x^x(T)+2\rho_y^y(T)+\rho_z^z(T)
\\
&    -\rho_{-{z}}^z(T)
    -\rho_z^x(T)-\rho_z^y(T)-\rho_{-{z}}^x(T)-\rho_{-{z}}^y(T)\biggr],}
\ee
where $\rho_j^k=\Tr\{\sigma^k\rho_j(T)\}$ for initial condition $\rho_{\pm j}(0)=(\sigma^I\pm\sigma^j)/2$.  As discussed in Sec.~\ref{quantum_bath}, the unitary evolution of this model assures $\rho^j_z = -\rho^j_{-z}$, simplifying the fidelity expression to
\be
\chi_{II}(T) = \frac{1+\rho_x^x(T)+\rho_y^y(T)+\rho_z^z(T)}{4}.
\ee
As a result, the infidelity is simply written
\be
\fl
\infid = \frac{\epsilon^2}{4}\int_0^T dt \int_0^t d\tau \int \frac{d\omega}{2\pi}
        \cos[\omega(t-\tau)]S(\omega)\Tr\{\mathbf{\Gamma}(t,\tau)\}+O(\epsilon^4),
\ee
and our generalized filter function is therefore simply
\be
F(z) = \frac{z^2}{4}\int_0^1 du \int_0^u dv \cos[z(u-v)]\Tr\{\mathbf{\Gamma}(Tu,Tv)\}.
\ee
Upon taking the trace, the matrix $\Gamma^{jk}(t,\tau)$ simplifies substantially to
\begin{equation}
\Tr[\mathbf{\Gamma}(t,\tau)]
    =8\{\cos[\magcon(t)-\magcon(\tau)]
    +[\dircon^z]^2[1-\cos[\magcon(t)-\magcon(\tau)]]\}.
\end{equation}
In this paper, we only consider control fields for which $\dircon^z(t)=0$ at all times.  In this case our general expression for $F(z)$ is
\be
F(z) = 2z^2\int_0^1 du \int_0^u dv \cos[z(u-v)]\cos[\magcon(Tu)-\magcon(Tv)].
\ee
This easily simplifies to \refeq{fftheory} in Sec.~\ref{filter_function}. \section*{References}

\end{document}